# MODEL PREDICTIVE CONTROL OF WIND TURBINES: A MODEL-ORIENTED REVIEW OF PREDICTION MODELS, CONTROL OBJECTIVES, AND NONLINEAR AERODYNAMIC FORMULATIONS

**Roya Khalili-Amirabadi[1], Mohsen Jalaeian-Farimani*[2], and Mahdi Ghane[1]**

*[1]Department of Computer Science, Electrical Engineering and Mathematical Sciences, Western Norway University of Applied Sciences*

*[2]Department of Electronics, Information and Bioengineering (DEIB), Politecnico di Milano, Milan, Italy*

** E-mails: roya.khalili.a@gmail.com; mohsen.jalaeian@polimi.it (Corresponding Author); mahdi.ghane@hvl.no*

**ABSTRACT**

Model predictive control (MPC) has been applied to wind-turbine regulation with markedly different internal prediction models, ranging from linearized low-order turbine models to nonlinear, aeroelastic, floating, and data-driven predictors. This review organizes the literature primarily by the model propagated by the controller and treats model order, predictor representation, MPC formulation, control objective, and validation plant as separate descriptors. Primary studies show that low-order linear or piecewise-affine predictors are attractive for constrained power and speed regulation, whereas flexible drivetrain, structural, and floating models are needed when drivetrain twist, tower motion, blade loads, or platform dynamics are explicit control quantities. Nonlinear and economic MPC studies further demonstrate coordinated pitch/torque control across broad operating regions, while preview, repetitive, and data-driven formulations address wind prediction, periodic loads, and model mismatch. Particular attention is given to aerodynamic pitch dependence: variable-pitch turbine dynamics can be non-affine with respect to applied pitch, and the term "NMPC" alone does not prove that the original nonlinear pitch map is retained inside the optimizer. The review therefore adopts a conservative, equation-based classification for direct nonlinear applied-pitch propagation. Here, "direct nonlinear applied-pitch propagation" means that the applied pitch at each prediction step is evaluated inside the original nonlinear aerodynamic map, rather than replacing that dependence by a linearized, scheduled, piecewise-affine, or learned surrogate. The literature supports a remaining integration challenge at the intersection of flexible drivetrain prediction, nonlinear aerodynamic pitch treatment, explicit constraints and load objectives, and validated adaptive or learned model correction.

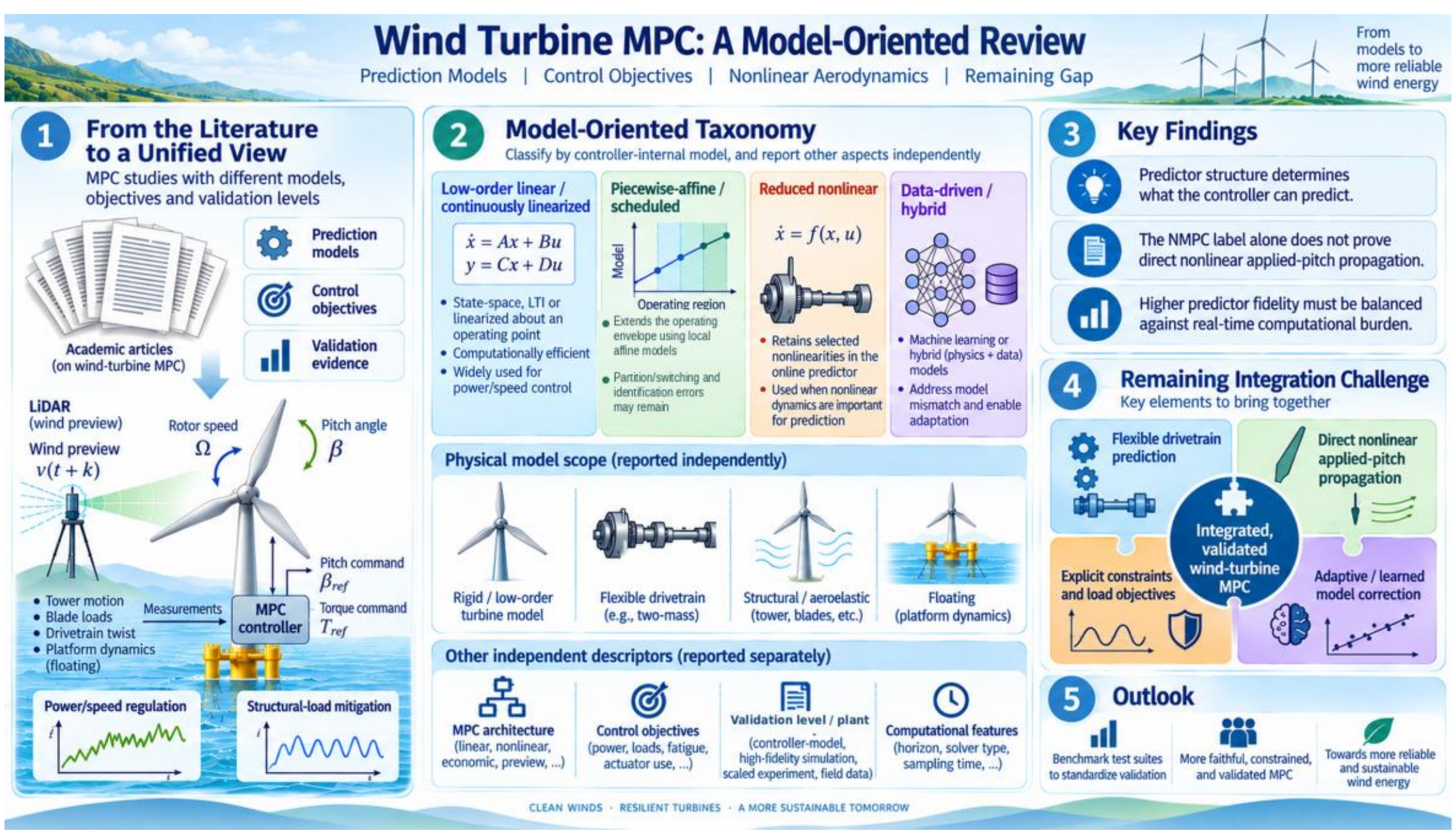

**Graphical abstract**

## 1. INTRODUCTION

Wind-turbine control is intrinsically multivariable and constrained: pitch and generator torque affect power capture, rotor speed, actuator use, drivetrain motion, and structural loading. MPC has therefore attracted sustained attention because it solves a finite-horizon constrained optimization problem and can coordinate competing objectives in a single controller [Mahmoud and Oyedeji, 2019, Henriksen et al., 2012, Jain et al., 2015]. The literature includes linear and continuously linearized MPC, nonlinear MPC (NMPC), economic NMPC, preview-enabled MPC, individual-pitch MPC, repetitive predictive control, floating-turbine MPC, and data-driven predictive control [Gros et al., 2013, Gros and Schild, 2017, Schlipf et al., 2013, Raach et al., 2014, Liu et al., 2021, Kayedpour et al., 2022]. Recent studies have further extended this landscape toward real-time and adaptive implementations, including parallel NMPC across the full operating range, time-varying economic NMPC with explicit drivetrain- and tower-fatigue objectives, and LPV-based adaptive MPC for pitch regulation and load reduction [Soleymani et al., 2023, Soleymani et al., 2024, Wen et al., 2026]. More recent developments have combined LiDAR-informed reinforcement learning with constrained NMPC cost adaptation and have integrated frequency-support MPC with explicit drivetrain-fatigue mitigation and field validation [Jin and Fu, 2026, Yang et al., 2026].

Existing reviews have generally organized the field by control strategy or application level rather than by the exact model propagated online. Lio et al. [Lio et al., 2014] surveyed applications and control benefits of MPC for wind turbines; Mahmoud and Oyedeji [Mahmoud and Oyedeji, 2019] reviewed adaptive and predictive wind-turbine control together with control-oriented modeling; and Lu et al. [Lu et al., 2024] organized MPC for large-scale wind-power integration across turbine, farm, cluster, and grid levels and multiple objectives. The present review differs by using controller-internal model structure as the primary classification axis and by recording predictor representation, MPC architecture, control objective, and validation plant independently. This separation resolves article-level distinctions, such as linearized versus direct nonlinear applied-pitch dynamics and controller predictor versus validation simulator, that are not determined by the MPC label alone.

A central difficulty in comparing these studies is that the label applied to the optimizer does not uniquely determine what the controller can predict. Henriksen et al. used MPC to handle hard and soft wind-turbine constraints over partial- and full-load operation [Henriksen et al., 2012]. Jain et al. used a continuously linearized turbine model and tuned the controller against power variation, pitch usage, tower displacement, drivetrain twist, and power-limit violation [Jain et al., 2015]. Schlipf et al. instead propagated a reduced nonlinear internal model with LiDAR wind preview and validated the controller against a much more detailed aeroelastic plant [Schlipf et al., 2013]. Evans and Lio later demonstrated that high-dimensional linear models obtained directly from aeroelastic software can be used in MPC while reducing online optimization degrees of freedom [Evans and Lio, 2022]. These examples show why prediction-model structure and validation-model fidelity must be reported separately. Broader learning-based control research further motivates this distinction: recent work has systematized reinforcement-learning-MPC integration for adaptive constrained control, investigated online learning for nonlinear systems with uncertain dynamics, and developed safe reinforcement-learning mechanisms with explicit state and input constraint enforcement [Jalaeian Farimani et al., 2026, Khalili Amirabadi et al., 2026, Khalili-Amirabadi et al., 2026].

Related nonlinear-control literature also illustrates resilience and estimation mechanisms outside the wind-turbine MPC domain. Adaptive fuzzy decentralized control has been developed for nonlinear large-scale cyber-physical systems under denial-of-service attacks [Zhao et al., 2024], while fuzzy adaptive bilateral cooperative control has been investigated for nonlinear multi agent systems with state/output quantization and observer-based state estimation [Zhao et al., 2025]. These studies are not classified as wind-turbine MPC evidence here; they are cited as broader examples of nonlinear approximation, observer design, communication constraints, and resilience concepts that can inform future cyber-resilient or distributed predictive-control architectures.

A second comparison issue is the aerodynamic pitch channel. The power and torque produced by a variable-speed, variable-pitch turbine depend nonlinearly on the tip-speed ratio and blade pitch. Bagheri and Sun explicitly formulated variable-speed, variable-pitch turbine dynamics as non-affine with respect to pitch angle [Bagheri and Sun, 2016]; Kong et al. likewise described the variable-speed turbine as a nonlinear, coupled, non-affine system in their nonlinear economic MPC formulation [Kong et al., 2020]. Consequently, calling a controller "nonlinear MPC" is insufficient to establish that the original nonlinear pitch dependence is propagated directly by the online optimizer.

The contribution of this review is to synthesize the supplied wind-turbine MPC literature with a model-oriented classification grounded in article-level evidence. The review separates (i) physical prediction-model structure, (ii) predictor representation, (iii) MPC formulation and sensing/actuation architecture, (iv) control objective, and (v) validation level. Detailed claims about a paper's states, constraints, or pitch treatment are made only when supported by the article or an authoritative publication record; otherwise the claim is not promoted to a more specific category.

## 2. REVIEW BASIS AND CLASSIFICATION RULES

The review focuses on turbine-level predictive control in which the online predictor contains aerodynamic, mechanical, drivetrain, structural, or floating-turbine dynamics. Converter-switching MPC without turbine mechanical states is outside the scope. Wind-farm supervisory MPC is included only where it provides evidence on the turbine predictor used for active-power allocation, as in the clustering-based and piecewise-affine models of Zhao et al. [Zhao et al., 2015a, Zhao et al., 2015b].

For clarity, three model entities are distinguished throughout this review. The nonlinear physical plant is the actual wind-energy-conversion system whose aerodynamic, drivetrain, structural, actuator, and environmental dynamics generate the physical response. A high-fidelity nonlinear validation simulator is a numerical surrogate used to assess the controller and may contain substantially more states and physics than the control model. The online predictor is the model propagated inside the MPC optimization at each control update [Schlipf et al., 2013, Evans and Lio, 2022]. Consequently, the use of a high-fidelity validation simulator does not imply that the same model is propagated online.

Five descriptors are used independently. *Physical model structure* distinguishes rigid/low-order turbine models, flexible drivetrain models, structurally augmented or aeroelastic models, floating models, and data-driven/hybrid predictors. *Predictor representation* records whether the online model is linearized, continuously linearized, scheduled, piecewise affine, reduced nonlinear, high-dimensional linearized aeroelastic, or learned. *MPC formulation* records linear MPC, NMPC, economic MPC/NMPC, repetitive or fault-tolerant predictive control, and related architectures. *Control objective* records the quantities optimized or constrained. *Validation level* distinguishes evaluation on the controller model from evaluation on a higher-fidelity simulator, scaled experiment, or field data. This separation follows the diversity visible in primary studies rather than treating all MPC implementations as instances of a single model class [Henriksen et al., 2012, Jain et al., 2015, Schlipf et al., 2013, Evans and Lio, 2022].

For aerodynamic pitch treatment, this review uses a deliberately conservative criterion. A study is described as retaining *direct nonlinear applied-pitch propagation* only when article-level equations or an explicit model description establish that the evolving applied pitch remains inside the nonlinear aerodynamic map used over the prediction horizon. Linearization, gain scheduling, piecewise-affine approximation, or Hammerstein/neural approximation are reported as their own predictor classes, even when they approximate an underlying nonlinear turbine [Zhao et al., 2015a, Zhao et al., 2015b, Kayedpour et al., 2022].

Source type is treated as an evidence-strength descriptor rather than as an exclusion rule. Peer-reviewed journal articles are given the greatest weight for detailed classification claims because they normally provide an archival, revised description and a fuller validation record. Peer-reviewed conference papers are retained when they contain the primary formulation or experimental result, whereas preprints are treated as provisional and are not used alone to support a strong article-level classification when a peer-reviewed version is available. Recent non-journal sources are therefore interpreted conservatively and cross-checked against equations, full-text descriptions, or authoritative publication records.

**Table 1:** Model-oriented classification used in this review.

| Descriptor | Representative classes | Question answered |
|---|---|---|
| Physical predictor | Low-order/rigid; flexible drivetrain; structural/aeroelastic; floating; data-driven/hybrid | Which physical quantities can be predicted directly? |
| Representation | Linearized; continuously linearized; PWA; reduced nonlinear; high-dimensional linearized aeroelastic; learned/Hammerstein | How are nonlinear and structural dynamics represented online? |
| MPC architecture | Linear MPC; NMPC; economic MPC/NMPC; preview; IPC; repetitive; fault-tolerant; distributed/data-driven | How is the finite-horizon control problem organized? |
| Objective | Power/speed; drivetrain; blade/tower loads; platform motion; fatigue; grid support | Which quantities are optimized or constrained? |
| Validation | Controller-model simulation; high-fidelity simulation; scaled experiment; field/reference data | How strongly is plant-model mismatch exposed? |

## 3. PREDICTION MODELS USED IN WIND-TURBINE MPC

### 3.1 Aerodynamic input structure

A common control-oriented description of aerodynamic power uses the power coefficient $C_p$ as a function of tip-speed ratio and applied blade pitch. In notation consistent with the reviewed nonlinear wind-turbine studies,

$$P_a = 1/2\, \rho A C_p(\lambda, \beta) v^3, \qquad \lambda = \frac{R\omega_r}{v}, \tag{1}$$

and aerodynamic torque is related to power by $T_a = P_a/\omega_r$ away from standstill. The control significance is not the algebraic form alone, but that $C_p$ depends on the applied pitch $\beta$. Bagheri and Sun identify pitch as a non-affine control input in the turbine dynamics, while Kong et al. explicitly formulate a nonlinear non-affine variable-speed wind-turbine model for economic MPC [Bagheri and Sun, 2016, Kong et al., 2020]. For this reason, model order and pitch non-affinity are treated as independent attributes.

When a paper replaces the global nonlinear dependence by local linear models, the resulting predictor should be reported as linearized or scheduled rather than as direct physical pitch non-affinity. This distinction is illustrated by continuously linearized MPC [Jain et al., 2015], by the model-uncertainty study of Odgaard and Hovgaard, which explicitly analyzes errors induced by linearization [Odgaard and Hovgaard, 2015], and by Zhao et al.'s clustering-based piecewise-affine turbine model [Zhao et al., 2015a, Zhao et al., 2015b].

### 3.2 Low-order and flexible drivetrain predictors

Low-order models are attractive when the main controlled variables are power, speed, and actuator activity. Their small state dimension facilitates frequent re-optimization and broad operating-envelope scheduling. Zhao et al. constructed a clustering-based wind-turbine model and then a piecewise-affine realization for distributed wind-farm active-power control, with power tracking and mechanical-load objectives [Zhao et al., 2015a, Zhao et al., 2015b]. Such models are useful supervisory predictors but should not be interpreted as direct nonlinear aerodynamic propagation because the nonlinear turbine is represented by affine local dynamics.

Flexible drivetrain modeling becomes important when drivetrain twist or related load metrics are explicit parts of the controller objective. Jain et al. include drivetrain twist among the performance measures used to tune linear MPC [Jain et al., 2015]. Henriksen et al. use a constrained turbine model able to operate across partial and full load regions and emphasize constraint handling in a flexible control-oriented representation [Henriksen et al., 2012]. Kong et al. include gearbox/drive-shaft dynamics together with tower dynamics in a nonlinear economic MPC model and penalize drive-shaft torsion as part of the economic objective [Kong et al., 2020]. These studies support the practical distinction between low-order power/speed predictors and flexible predictors that make drivetrain motion an explicit optimization quantity.

### 3.3 Structural and aeroelastic predictors

Structural states can enter MPC either through a reduced control model or through a high-dimensional linearization of a detailed aeroelastic model. Jain et al. include tower displacement and drivetrain twist in the multiobjective tuning problem [Jain et al., 2015]. Schlipf et al. use a reduced nonlinear model inside the NMPC but test it using a detailed FAST-based aeroelastic model; the paper explicitly separates the controller's reduced model from the higher-order simulation plant [Schlipf et al., 2013]. Evans and Lio take a different route: they use a detailed linearized model exported from an aeroelastic simulation package, retain a large state space, and reduce online computational burden mainly by parameterizing the control trajectory with few optimization variables [Evans and Lio, 2022].

Individual-pitch control (IPC) further increases actuation dimension because the controller computes blade-specific pitch commands. Mirzaei et al. formulate MPC-based IPC using LIDAR information and linearize the nonlinear turbine model at operating points scheduled by wind speed [Mirzaei et al., 2013]. Sinner et al. experimentally test a preview-enabled pitch MPC on a scaled turbine, providing physical evidence for the use of preview information in constrained wind-turbine control [Sinner et al., 2022].

### 3.4 Floating offshore wind-turbine predictors

Floating turbines add platform motion and wave-induced dynamics to the wind-turbine control problem. Raach et al. present NMPC with individual pitch control using a reduced nonlinear floating-wind-turbine model with wind and wave preview and generator-torque coordination [Raach et al., 2014]. Thiery et al. address failure compensation using individual blade pitch within an MPC framework for a floating turbine [Thiery et al., 2015]. These works show that "floating MPC" is defined by the platform-coupled prediction problem; the drivetrain order must still be identified separately rather than inferred from the floating label.

High-fidelity floating reference studies also clarify the role of validation. Full-scale Hywind Demo measurements have been compared with simulations for platform motions, tower bending, mooring loads, power, rotor speed, and pitch [Skaare et al., 2015]. OC5 compared FAST-based simulations of the DeepCwind floating system with wave-tank data [Wendt et al., 2019], while OC6 Phase III validated aerodynamic loading under prescribed floating-platform motion using wind-tunnel measurements [Bergua et al., 2023]. Such studies provide validation environments and physical benchmarks; they are not evidence that an MPC necessarily propagates the same high-order model online.

### 3.5 Data-driven and hybrid predictors

Data-driven predictive control replaces or augments first-principles dynamics with identified input-output models. Kayedpour et al. use a cascaded Hammerstein model in which neural networks approximate the nonlinear steady-state aerodynamic behavior and an ARX model represents the linear dynamic part; the resulting MPC coordinates pitch and torque for frequency containment reserve while avoiding a fully nonlinear online optimization [Kayedpour et al., 2022]. Liu et al. use a fully data-driven subspace predictive repetitive-control framework with explicit blade-pitch angle and rate constraints to reject periodic loads on floating offshore wind turbines [Liu et al., 2021].

More recent work moves from fixed learned models toward adaptive correction of a reduced-order model. Anand and Bottasso formulate adaptive economic wind-turbine control in which a correction term is learned to reduce mismatch between the MPC reduced-order model and the turbine response, while economic and fatigue-related objectives remain inside the predictive-control framework [Anand and Bottasso, 2026]. This study provides direct article-level evidence that model adaptation can be combined with economic predictive control; it should not, however, be conflated with the Hammerstein architecture of Kayedpour et al., because the location and role of learning in the control loop differ [Kayedpour et al., 2022, Anand and Bottasso, 2026].

## 4. MPC FORMULATIONS, OBJECTIVES, AND VALIDATION EVIDENCE

The prediction model and the optimization formulation are separate design choices. Linear MPC remains attractive when a sufficiently accurate linearized model is available and computational efficiency is important. Jain et al. demonstrate systematic tuning of continuously linearized MPC across several competing performance indices, while Odgaard and Hovgaard show that linearization error can materially degrade performance in nonlinear operating regions and propose parameter adjustment of the linearized model [Jain et al., 2015, Odgaard and Hovgaard, 2015].

NMPC permits direct propagation of a nonlinear control model. Gros et al. combine moving-horizon estimation with real-time NMPC for wind-turbine control [Gros et al., 2013]. Schlipf et al. use nonlinear prediction with LiDAR preview to reduce extreme and fatigue loads in simulation against a detailed aeroelastic plant [Schlipf et al., 2013]. Raach et al. extend NMPC to floating-turbine IPC [Raach et al., 2014]. These papers demonstrate nonlinear predictive-control architectures, but the specific nonlinearities retained by the internal model differ and therefore must be described from each article rather than inferred from the common NMPC label.

Economic MPC changes the objective from primarily reference tracking toward a direct operational criterion. Gros and Schild compare economic NMPC with classical NMPC and study robustness and wind-preview limitations [Gros and Schild, 2017]. Kong et al. construct a nonlinear economic objective balancing generated power with drive-shaft torsion, tower fore-aft motion, and actuator-related terms while imposing pitch, torque, rate, speed, and electrical-power constraints [Kong et al., 2020]. Loew and Bottasso incorporate fatigue through online rainflow counting and stress history in the MPC cost, with moving-horizon state estimation and LiDAR-based wind prediction [Loew and Bottasso, 2022]. These studies show that "economic" and "nonlinear" are independent descriptors: an economic objective does not, by itself, identify the predictor representation.

Preview is likewise a sensing/optimization architecture rather than a model-order class. Schlipf et al. place LiDAR-derived wind information into the nonlinear prediction horizon [Schlipf et al., 2013]; Mirzaei et al. use uncertain LiDAR measurements in IPC-MPC [Mirzaei et al., 2013]; and Sinner et al. provide scaled experimental validation of preview-enabled blade-pitch MPC [Sinner et al., 2022]. The repeated appearance of preview across linearized and nonlinear formulations illustrates why disturbance information should be classified separately from the physical prediction model.

Fault-tolerant MPC forms a distinct subdomain in which the predictive controller modifies future inputs, constraints, or model usage in response to detected actuator, sensor, or component faults. Representative studies illustrate different prediction-model structures. Thiery et al. employ individual-pitch MPC with failure compensation for a floating wind turbine [Thiery et al., 2015]; Jain and Yame formulate fault-tolerant economic MPC using a control-oriented turbine model that includes drivetrain and tower dynamics together with real-time constraint reconfiguration [Jain and Yame, 2019]; and Feng and Patton combine MPC with a Takagi–Sugeno fuzzy model and observer-based fault estimation [Feng and Patton, 2014]. Thus, fault tolerance should be treated as an MPC architecture or function rather than as a unique prediction-model class; the predictor structure must still be identified separately from the underlying article.

**Table 2:** Representative article-level evidence used in the model-oriented synthesis. “Pitch treatment” refers to the representation established by the cited article, not by the controller label alone. The final column reports horizon, solver type, and sampling time when these were explicitly verified; NR denotes not reported or not verified at article level in the present evidence set.

| Study | Predictor/model | Main actuation or information | Main objective/evidence | Pitch treatment / validation | Core computational features |
|---|---|---|---|---|---|
| Henriksen et al. [Henriksen et al., 2012] | Constrained control-oriented turbine model | Pitch/torque within hybrid operating-region control | Hard/soft constraints; partial/full-load operation | Model-based MPC; higher-fidelity evaluation reported in the literature | Solver/horizon/Ts: NR |
| Jain et al. [Jain et al., 2015] | Continuously linearized turbine model | Turbine control inputs | Power variation, pitch use, tower displacement, drivetrain twist | Linearized predictor; nonlinear simulation model for evaluation | Gurobi convex optimization; horizon 2 s; Ts 0.1 s |
| Gros et al. [Gros et al., 2013] | Nonlinear control model + MHE | Pitch/torque | Real-time NMPC and estimation | Nonlinear predictor; real-time-oriented implementation | ACADO RTI; condensed QPs; horizon 10 s; Ts 0.2 s |
| Schlipf et al. [Schlipf et al., 2013] | Reduced nonlinear internal model | Collective pitch, electrical torque, LiDAR preview | Extreme/fatigue-load reduction | Reduced nonlinear model; detailed FAST aeroelastic validation | SQP/HQP (Omuses); horizon 10 s; update 0.2 s |
| Raach et al. [Raach et al., 2014] | Reduced nonlinear floating model | Individual pitch, generator torque, wind/wave preview | Floating power/load regulation | Nonlinear floating predictor; simulation study | Solver/horizon/Ts: NR |
| Zhao et al. [Zhao et al., 2015a, Zhao et al., 2015b] | Clustering-based PWA turbine model | Active-power references | Wind-farm power tracking and mechanical-load minimization | Piecewise-affine predictor | Fast-gradient method via dual decomposition; horizon/Ts: NR |
| Kong et al. [Kong et al., 2020] | Nonlinear gearbox + tower model | Pitch and generator torque | Power maximization with drive-shaft/tower-load mitigation | Nonlinear non-affine model; FAST benchmark also reported | Solver/horizon/Ts: NR |
| Sinner et al. [Sinner et al., 2022] | Preview-enabled reduced control model | Blade pitch + preview | Preview pitch control | Scaled wind-tunnel experiment | Solver/horizon/Ts: NR |
| Kayedpour et al. [Kayedpour et al., 2022] | Cascaded Hammerstein neural + ARX | Pitch and torque | Frequency containment reserve | Learned nonlinear static map + linear dynamic model | Online linearization + QP; horizon/Ts: NR |
| Liu et al. [Liu et al., 2021] | Data-driven subspace repetitive predictor | Individual pitch | Periodic-load rejection under pitch constraints | Fully data-driven predictive repetitive control | Solver/horizon/Ts: NR |
| Loew and Bottasso [Loew and | Reduced turbine model + MHE + LiDAR | Pitch with predicted wind | Fatigue-aware MPC via online rainflow counting | Model mismatch deliberately | 1 QP/step; horizons 1, 2, 4, 8 s; Ts 0.1 s |

| Study | Predictor/model | Main actuation or information | Main objective/evidence | Pitch treatment / validation | Core computational features |
|---|---|---|---|---|---|
| Bottasso, 2022] | | | | included in simulation | |
| Anand and Bottasso [Anand and Bottasso, 2026] | Adaptive corrected ROM | Predictive turbine control | Economic/fatigue objective with model adaptation | Learned correction of controller-internal ROM | Acados SQP + HPIPM QP; horizon 2 s; Ts 0.1 s |

**Table 3:** Cross-category comparison of representative online predictor classes. Computational footprint denotes the typical online optimization structure and does not imply identical hardware or horizon settings across studies.

| Predictor class | Principal merits | Principal limitations | Typical online computational footprint | Operational applicability |
|---|---|---|---|---|
| Low-order linear / continuously linearized | Transparent; fast; constraints handled with structured convex optimization; straightforward relinearization. | Local validity; can omit direct pitch non-affinity and structural/floating modes. | Low: typically a small/medium QP; high update rates are feasible. | Power/speed regulation, actuator constraints, and local operating regions. |
| Piecewise-affine / scheduled | Extends the operating envelope with local affine models while retaining structured optimization. | Partition/switching and identification errors; remains a surrogate of global nonlinearity. | Low–moderate: QP/fast-gradient solution plus mode or region handling. | Multi-region regulation, farm-level allocation, and supervisory control. |
| Reduced nonlinear | Retains selected aerodynamic, drivetrain, structural, or floating nonlinearities and preview dynamics. | Nonconvex NLP; convergence/warm-start sensitivity; model-reduction error. | Moderate–high: nonlinear simulation with NLP/SQP/RTI iterations. | Broad-envelope power/load control, preview control, and floating turbines. |
| Data-driven / hybrid | Captures model mismatch from data; structured hybrid forms can retain efficient online optimization. | Requires representative data; extrapolation, interpretability, robustness, and adaptation overhead must be managed. | Variable: QP for linear/Hammerstein forms up to NLP for nonlinear learned predictors or adaptive schemes. | Repetitive-load control, frequency support, model correction, and adaptation. |

## 5. DIRECT NONLINEAR APPLIED-PITCH DEPENDENCE

The reviewed literature supports a strict distinction between a nonlinear turbine model and a model that is non-affine in a particular control channel. Bagheri and Sun state explicitly that the turbine dynamics are non-affine with respect to pitch angle as a control input and design a nonlinear adaptive controller for that structure [Bagheri and Sun, 2016]. Bagheri, Behjat, and Sun later address a related non-affine pitch-control problem using radial-basis-function neural networks and a transformation to a control-affine form [Bagheri et al., 2022]. Kong et al. also describe the variable-speed wind-turbine system used for nonlinear economic MPC as nonlinear, strongly coupled, and non-affine [Kong et al., 2020].

For MPC classification, however, the decisive issue is the model actually propagated online. A continuously linearized model [Jain et al., 2015], a PWA turbine model [Zhao et al., 2015a, Zhao et al., 2015b], and a Hammerstein/neural approximation [Kayedpour et al., 2022] may all originate from nonlinear turbine physics but use different online input representations. This review therefore does not assign "direct nonlinear applied-pitch propagation" from the word NMPC, from the presence of nonlinear plant dynamics in a validation simulator, or from a learned nonlinear surrogate alone.

This conservative rule also prevents a common validation error: a controller may be tested on a detailed aeroelastic or floating model without using that model inside the optimizer. Schlipf et al. explicitly state that the full aeroelastic model is too complex for their NMPC and use a reduced model internally [Schlipf et al., 2013]. Evans and Lio, conversely, deliberately retain a high-dimensional linearized aeroelastic state model while reducing the number of online decision variables [Evans and Lio, 2022]. Reporting "controller predictor" and "validation plant" as separate fields is therefore necessary for meaningful comparisons.

## 6. VALIDATION, MODEL REDUCTION, AND COMPUTATIONAL TRACTABILITY

Validation quality differs substantially across the literature. Some studies evaluate MPC on nonlinear simulation models distinct from the controller's linearized predictor [Jain et al., 2015, Odgaard and Hovgaard, 2015]. Schlipf et al. evaluate a reduced nonlinear NMPC model against a detailed aeroelastic FAST plant [Schlipf et al., 2013]. Sinner et al. extend the evidence chain to a scaled wind-tunnel experiment [Sinner et al., 2022]. Floating-wind reference studies provide still broader validation context: Hywind Demo offers full-scale measurement/simulation comparison [Skaare et al., 2015], OC5 compares FAST-model results with wave-tank data [Wendt et al., 2019], and OC6 validates floating-platform aerodynamic loading against wind-tunnel measurements [Bergua et al., 2023].

High-fidelity modeling is useful even when it is not propagated online. BeamDyn provides a high-fidelity structural blade solver within the FAST modular framework [Wang et al., 2017], and Yang et al. develop an aero-hydro-servo-elastic coupling framework for nonlinear floating-wind-turbine analysis and compare it with OpenFAST [Yang et al., 2020]. These tools are relevant as validation or model-development environments, but an MPC study should not be labeled "high-order online MPC" solely because such a plant is used for testing.

Computational strategies also differ. Gros et al. target real-time NMPC together with MHE [Gros et al., 2013]; Evans and Lio retain many model states but reduce optimization degrees of freedom using trajectory parameterization [Evans and Lio, 2022]; Kayedpour et al. avoid nonlinear online optimization by combining a Hammerstein representation with online linearization and quadratic programming [Kayedpour et al., 2022]. Accordingly, "model fidelity" cannot be reduced to state count alone: the online computational burden depends on predictor form, estimation, decision-variable parameterization, horizon, and solver structure.

At a practical level, predictor fidelity and real-time tractability must be balanced. Higher-fidelity models can improve prediction of loads, constraints, and cross-coupled dynamics, but they increase online cost through nonlinear integration, more decision variables, denser sensitivities, estimation, and stricter convergence requirements. Linear and piecewise-affine models usually admit QP or other structured convex solvers and therefore favor high update rates, at the cost of local-model or partition error. Reduced nonlinear predictors retain selected physical nonlinearities but typically require NLP/SQP/RTI iterations and effective warm starts. Data-driven and Hammerstein/hybrid predictors can shift part of the complexity to offline identification and may retain a QP online, whereas online adaptation or nonlinear learned correction adds evaluation and robustness overhead. The appropriate fidelity is therefore the minimum model complexity that reliably captures the controlled quantities and active constraints within the available sample time.

Performance metrics themselves also limit cross-study comparison. Mean or RMS power and speed errors can hide short constraint violations and extreme events; aggregate pitch activity does not uniquely quantify actuator fatigue; tower or drivetrain RMS measures are not interchangeable with damage-equivalent loads or rainflow-based fatigue; and energy/load improvements depend strongly on wind realization, turbulence intensity, operating region, preview quality, controller horizon, baseline tuning, and validation plant. Strong comparisons should report distributions and extreme/constraint metrics, energy capture, actuator usage, structural fatigue measures, identical disturbance cases or statistical ensembles, and computational deadline performance in addition to single aggregate indices.

## 7. RESEARCH GAPS AND OUTLOOK

The primary studies reviewed here already establish most individual ingredients that might otherwise be presented as novel. Flexible and load-aware MPC is present in constrained and continuously linearized turbine-control studies [Henriksen et al., 2012, Jain et al., 2015, Kong et al., 2020]. Real-time NMPC and economic NMPC are established [Gros et al., 2013, Gros and Schild, 2017, Kong et al., 2020]. LiDAR preview and experimental preview control are established [Schlipf et al., 2013, Mirzaei et al., 2013, Sinner et al., 2022]. Floating NMPC and constrained repetitive floating control are established [Raach et al., 2014, Liu et al., 2021]. Learned and adaptive predictive models are established through Hammerstein/neural MPC and adaptive economic control [Kayedpour et al., 2022, Anand and Bottasso, 2026]. Explicit non-affine wind-turbine control also predates the present review outside MPC [Bagheri and Sun, 2016, Bagheri et al., 2022].

The defensible gap is therefore integrative rather than component-wise. Across the verified studies in this review, the strongest remaining direction is to combine a flexible drivetrain predictor with direct nonlinear applied-pitch aerodynamic propagation, explicit pitch/torque and drivetrain-load constraints or objectives, and a validated adaptive/model-correction mechanism, while preserving a clear separation between the controller-internal model and the high-fidelity validation plant. This statement is intentionally evidence-bounded: it identifies a sparsely represented combination in the verified corpus rather than claiming an absolute first.

The evidence supporting this integrative gap is drawn predominantly from utility-scale, megawatt-class onshore and offshore wind-turbine studies. Accordingly, the evidence-bounded gap identified here should be interpreted primarily within the utility-scale context represented by the verified corpus. Similar combinations of flexible drivetrain prediction, direct nonlinear applied-pitch aerodynamic propagation, explicit pitch/torque and drivetrain-load constraints or objectives, and adaptive model correction may also be relevant to variable-pitch small-scale turbines. However, the present review does not extend the literature-gap claim to small-scale wind turbines because that population was not systematically covered and may involve different scale-dependent dynamics, actuator bandwidths, sensing configurations, and computational constraints.

A second gap concerns reporting. Several comparisons become ambiguous when the paper title or abstract gives an MPC label but does not immediately identify online states, the representation of pitch, the validation plant, or whether learning modifies the predictor or the command-generation layer. Future wind-turbine MPC papers would be substantially easier to compare if they reported, at minimum, the prediction state vector, manipulated inputs, disturbance preview, input and state constraints, pitch representation, estimation method, validation plant, horizon, sampling time, solver, and measured computation time.

Future benchmark test suites could reduce these ambiguities by prescribing common reference turbines, wind/disturbance cases, actuator and constraint definitions, baseline controllers, and multiple validation tiers (controller-model, high-fidelity simulation, and, where available, experiment or field data). A benchmark should also require common reporting of performance metrics together with predictor states, horizon, sampling time, solver, hardware, measured computation time, and deadline violations. Identical disturbance seeds and openly specified evaluation scripts would make predictor-fidelity and computational trade-offs reproducible across MPC categories.

Finally, model adaptation should be evaluated against explicit plant-model mismatch and not only against the model used to generate training data. Anand and Bottasso directly motivate adaptive correction by the limitations of a reduced-order MPC model [Anand and Bottasso, 2026]; Loew and Bottasso explicitly introduce model mismatch, MHE, and realistic LiDAR processing in fatigue-aware MPC assessment [Loew and Bottasso, 2022]. These studies provide an article-based foundation for future work on adaptive predictive control with transparent uncertainty and constraint treatment.

## 8. CONCLUSIONS

Wind-turbine MPC cannot be compared reliably by controller label alone. Primary studies demonstrate low-order linearized and piecewise-affine models for efficient constrained control, flexible and structural models for drivetrain/tower objectives, reduced nonlinear models for LiDAR-assisted NMPC, floating models for platform-coupled control, and learned predictors for data-enhanced operation. The online predictor, optimization formulation, controlled objective, and validation plant should therefore be classified independently.

The reviewed articles also support a careful treatment of pitch nonlinearity. Variable-pitch turbine dynamics can be non-affine in the pitch channel, but "NMPC" does not by itself prove that the original nonlinear aerodynamic pitch map is retained by the optimizer. A conservative equation-based classification avoids conflating nonlinear plants, linearized predictors, piecewise-affine approximations, and learned surrogates.

Within the verified literature, the most defensible research opportunity is the integrated use of flexible drivetrain prediction, direct nonlinear applied-pitch aerodynamics, explicit pitch/torque and drivetrain-load constraints or objectives, and validated adaptive model correction. The value of such a contribution would lie in the verified coexistence of these features and in high-fidelity or experimental validation, not in any one ingredient considered in isolation.

**Author Contributions:** All authors contributed equally to the preparation of this manuscript.
**Conflict of Interest:** The authors declare no conflict of interest.
**Data Availability:** No code or datasets were generated or analyzed in this study.